# The Implications of Network-Centric Software Systems on Software Architecture: A Critical Evaluation

Amine Chigani and James D. Arthur

Department of Computer Science

Virginia Polytechnic Institute and State University

Blacksburg, VA 24061

(703) 401-1586, (540) 231-7538

{achigani, arthur}@vt.edu

## ABSTRACT

The purpose of this paper is to evaluate the impact of emerging network-centric software systems on the field of software architecture. We first develop an insight concerning the term "network-centric" by presenting its origin and its implications within the context of software architecture. On the basis of this insight, we present our definition of a network-centric framework and its distinguishing characteristics. We then enumerate the challenges that face the field of software architecture as software development shifts from a platform-centric to a network-centric model. In order to face these challenges, we propose a formal approach embodied in a new architectural style that supports overcoming these challenges at the architectural level. Finally, we conclude by presenting an illustrative example to demonstrate the usefulness of the concepts of network centricity, summarizing our contributions, and linking our approach to future work that needs to be done in this area.

## Categories and Subject Descriptors

D.2.11 [**Software Engineering**]: Software Architectures – domain specific architectures, patterns

## 1. INTRODUCTION

Recent years have revealed a clear transformation from a platform-centric to a network-centric software development model. Conventional software use focuses on applications that are installed and updated manually [1, 4] on local devices such as PCs, mainframes, and others. Increasingly, the need to use other interface devices to access remote data and computational resources has become inevitable. Additionally, technological breakthroughs in hardware and network communications have opened the door for the software engineering community to address larger and more complex problems that were, until more recently, unsolvable. Furthermore, software acquisition has raised numerous challenges to integrate existing applications with the acquired ones. As a result, distributed, grid, service oriented computing, and other similar disciplines have emerged as products of the shift towards network-centric computing.

Amongst the critical issues in the design and implementation of any software-intensive system is its architecture [1, 10, 15]. Network-centric software systems are no different. They, too, require an underlying software architecture that ensures that the end-product satisfies key quality attributes such as performance, reliability, modifiability, and others. Additionally, because of its distributed nature and its intense reliance on networked communications, network-centric software systems require an underlying architecture that also ensures interoperability, flexibility, security, robust connectivity, and many other related characteristics.

In this paper, we critically evaluate the implications that emerging network-centric software systems have on the field of software architecture. In doing so, we provide a concrete characterization of network-centric software systems based on the initial needs that have driven this shift to a network-centric development paradigm. The following is the outline for the rest of the paper. Section 2 presents a background of the origin of network centricity as a term within the software engineering vocabulary. Section 3 describes the major aspects of the network-centric framework. Section 4 discusses several challenges facing software architecture as an engineering discipline with the introduction of network-centric software systems. Section 5 introduces our proposal that emerging network-centric software systems have led to the recognition of a new architectural style, which still needs to be officially formalized. Section 6 illustrates the potential value of this style by presenting a real-life system that fits our characterization of a network-centric software system. Section 7 includes our conclusions. Finally, section 8 suggests future research work that needs to be done in order to complete the documentation of this new architectural style, and establish its validity among other styles.

## 2. NETWORK-CENTRICITY: A BACKGROUND

"Network-centric" is a phrase that has been used loosely in many areas of the software engineering landscape. Among these areas is the field of software architecture [4, 13, 16, 20]. Understanding the origin of this term and its background enables us to use it more accurately to describe what it means in the context of software architecture.

The term "network-centric" originated from the DOD network-centric warfare (NCW), now commonly called network-centric operations (NCO). NCO is an emerging theory of war that seeks to translate an information advantage into a competitive warfighting advantage through the robust networking of well-informed, geographically-dispersed forces allowing new forms of warfighting organizational behavior. NCO's basic tenets include:

- Utilizing technological advantages to support warfighters in the battlefield

- Networking all systems used by US armed forces

- Achieving shared awareness of the battlefield amongst all members of the US armed forces [2]

To achieve its goals, NCO depends on many technologies including network architectures, satellites, radio bandwidth, unmanned vehicles, nanotechnology, processing power, and software systems. The last aspect (software systems) is what our research work focuses on: *What is different about this class of systems and how can we develop them?*

Many argue that the military has borrowed the term "network centricity" from existing business models that software corporations, such as Oracle, have developed to integrate its diverse and distributed assets. Others argue that the term originated from DOD and has found its way to industry as companies compete for government contracts to develop and provide tools, capabilities, and support mechanisms for NCO. Irrelevant to our discussion is whether the origin is NCO or industry; yet, it is critical that we understand the goals of network centricity within the context of warfighting, and within the context of software-intensive systems.

Similar to NCO, network-centric software systems are systems that focus substantially on their communications element. These systems promise to be the answer to today's most pressing computing challenge: Application and data integration [25]; that is, taking different applications running on different platforms (i.e., OSs), built with different object models, expressed using different programming languages [3], accessing different remote and local data repositories and integrating them into robust systems for supporting critical business processes or scientific research programs. The purpose of becoming network-centric is to be able to build software systems by integrating a mix of existing and new applications, and ensuring that the end-product is capable of integrating with other net-ready systems.

In fact, one of the distinctive characteristics of network-centric software systems is that their communicating elements are, to a large extent, loosely-coupled sub-systems that work together to solve a large and complex problem that cannot be solved by any individual element. The idea of a network-centric software system is that of a "system of systems" (Figure 1). Application developers increasingly want and need to reach beyond tightly-coupled client-server environments to access functionality on remote systems that are different in design and implementation, and which are possibly owned and managed by other organizations [3].

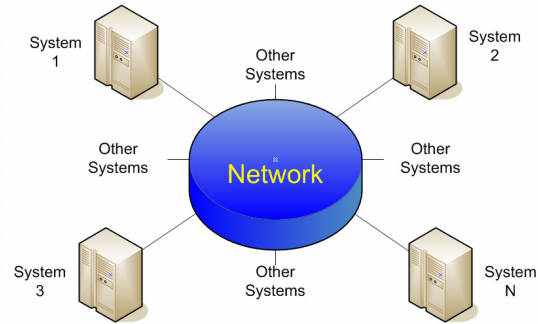

**Figure 1. The of notion of a "system of systems"**

In the following section, we discuss in detail the network-centric framework and its implications on the field of software architecture. This framework brings about challenges that affect software development, and in particular, software architecture, and that require solutions at the architecture level to face these challenges.

## 3. NETWORK-CENTRICITY: AN EMERGING SOFTWARE DEVELOPMENT FRAMEWORK

Network-centric computing embodies the "information age invasion" [1] of almost every area of science, art, business, education, government, and others. It represents a new way of thinking about how software engineers can accomplish their missions more efficiently and with increased capability by orders of magnitude. This view is supported by the accumulating software architectures and frameworks that have network-centric characteristics. Balci et al survey existing software architectures and frameworks that claim to address the network-centricity issue, and that are adopted by a number of government and industry institutions [2].

Within the network-centric development framework, the focus is on two main constituents: 1) the network and communication types among the software system's components, and 2) the software system's behavior at runtime. Further, there are two important aspects in the network-centric framework: A technology aspect and a human aspect. To elaborate, the network and communication types between the system's components correspond to the information technology side of the network-centric framework. Advances in networking technologies have been the drive that has led to the spread of a network-centric culture amongst software developers. On the other hand, the architects make the decisions on how elements of the system behave and communicate with one another in a networked setting to achieve a common objective. Therefore, the system behavior at runtime is driven by a human behavior manifested in the architects' design choices made during the software architecture design.

In our preliminary research, we have identified at least four characteristics that distinguish network-centric systems from other systems. A network-centric software system has:

- a *system of systems* perspective

- an underlying *networked configuration* that embodies the runtime environment on which the system's components interact and limits components' interaction to information exchange

- an *emergent, dynamic runtime behavior*, which means that the system's actual interacting components are not necessarily known until runtime

- a *fluid, dynamically-defined decentralized control*, which means that control over the system's functionality is not necessarily owned by a particular component; rather, this control changes based on which function the system is performing and which component has initiated the system's execution

The network-centricity concepts have several implications on software engineering, and in particular, the software design (architecture) phase. In traditional software systems, a common theme has been that they are constructed as closed systems managed by single organizations [1]. Although some components are usually reused or obtained from other internal and/or external applications, the entire system comes under the control of the designer or architect once integrated. For such systems, architectural design has often resulted in architectures that do not easily allow any dynamic behavior at runtime [1]. In this setting, architects must know beforehand where components will be located and how to interface with them. Architects have control over all components and therefore are able to predict their behavior at runtime.

However, this assumption is void within the network-centric model. A network-centric software system may have a central objective but may not necessarily have a centralized control. A leading application of the concepts of network-centricity is represented in the Internet [4]. The Internet is a collaborative network of networks that exhibits an emergent behavior that is a result of its complex architecture. Nevertheless, the Internet structure is facilitated by a minimal set of standards [1] in the form of protocols that describe how to exchange data over the network. These protocols are independent of the hardware or software applications that use the Internet. More importantly, adherence to these protocols is voluntary with no central authority that posses coercive power. Websites, web services, and other Internet-based activities are managed by their individual organizations and the decision to join and/or leave the Internet network is solely in the hands of that organization.

Architects of such systems face an emerging set of challenges. Their implications go beyond creating a single reference architecture that will support the design and implementation of many kinds of network-centric software systems. Rather, these challenges induce the need for a general approach to designing software architectures for this kind of systems. We believe that this approach must be in terms of a new architectural style, added to the reservoir of existing styles, which will enable architects to design systems that answer to both the demands of network centricity and their respective problem domain. The concept of a style defines the features of a family of software architectures for a particular class of systems [15, 24], network-centric software systems is one such class. The following section discusses the most relevant challenges that face the software architecture community in both research and industry.

# 4. RELEVANT CHALLENGES RELATED TO NETWORK-CENTRICITY

Many challenges that the software architecture community faces are not specific to network centricity. This is because the field of software architecture itself is in its nascent stage. Therefore, in this section we focus only on those challenges that are introduced by network-centric software systems. The following are the most dominant ones that we have identified by investigating software systems and software architectures that exhibit network-centric characteristics.

## 4.1 Standardization

The first challenge is the related need to develop software architectures that flexibly accommodate applications and services provided by various developers. An emerging trend in software development efforts is that systems are composed out of a mix of local and remote computing capabilities, requiring architectural support that accommodates interoperability, modifiability, connectivity, security, and other desirable operational qualities [6]. Thus, we argue that this support should come in the form of an architectural style that facilitates the generation of systems using a dynamically-formed coalition of distributed resources. More specifically, new standards (similar to Internet protocols) need to be established for building new components and making existing one net-ready.

## 4.2 Scalability

Building on the analogy between network-centric software systems and the Internet, a second challenge emerges: The need for scalable architectures that can evolve and that can handle component complexity and variability similar to architecture of the Internet. Network-centric software systems, being systems of systems, incorporate different components that require different architectural representations and various forms of communication. While many of the existing architectural styles will likely apply, the details of their application will need to change. Thus, we see that there is a need to define a new architectural style that accommodates these changes.

For instance, implicit invocation is a widely-accepted method of designing software systems. Implicit invocation is a style of software architecture in which a system is organized around event handling – broadcasting and subscribing to events. On one hand, this style allows for heterogeneous components to be integrated into systems that have low-coupling and high-cohesion with are two indispensable qualities of any software system. On the other hand, architects must make assumptions about certain qualities which are crucial to the system such as the reliability of event delivery and routing of messages. In a network-centric model, all such assumptions are uncertain [1]. Therefore, we are further convinced that we need novel techniques that allow architects to design these systems in such a way that it accommodates their dynamic growth.

## 4.3 On-Demand Composition

The third challenge is the need to develop architectures that permit end users to form their own system composition. With the rapid growth of the Internet, an increasing number of users are in a position to assemble and tailor services. Such users may have minimal technical expertise, and yet, will still want a sufficiently strong guarantee that the parts will work together in the ways they expect [1].

Architects must to find ways to support such needs for network-centric software systems. A network-centric architecture has to encompass characteristics that facilitate the generation of systems which are modifiable and that support an on-demand integration of new components.

## 4.4 Robust Connectivity

The fourth challenge that faces designers of network-centric software systems is the need for a robust infrastructure, which supports computing through a large number of independent, heterogeneous, distributed, dynamically-integrated components. For instance, the Internet infrastructure supports a broad range of resources such as primary information, communication mechanisms, web applications, services, and many others [1]. A common characteristic among these resources is independence – both operational and managerial. They can join and leave the network at will, they can invoke other resources and can be invoked, and, most importantly, they evolve independently of each other. Similarly, a network-centric software system must have an underlying infrastructure that facilitates a decentralized control over the system elements. Elements are selected and composed based on the task that needs to be carried out.

Due to the intrinsic complexity of automating the selection and composition process, architects must focus on the interface requirements between the elements of a network-centric software system. Within a network-centric model, architects do not necessarily have implementation knowledge about the components that are developed by other entities. In addition, the integration of incorporated components may be unfeasible if these components have static interface specifications. For instance, the integration of a component packaged to interact via remote procedure calls with a component packaged to interact via shared data can be a difficult task [1].

These are added challenges for architectural design. Creating an architectural style that facilitates the consideration of these challenges at the architecture level seems a reasonable proposition.

## 4.5 Security

Security models focus on the secure exchange of information among components of a system to meet the requirements defined by the problem domain. In the case of network-centric software systems, the intense reliance on networked communications brings about more security risks and concerns. Security cannot be an added feature to the system; it needs to be built into the system. Therefore, architectures that support the generation of network-centric software systems need to provide the capability to have security technologies built into the appropriate elements of the system infrastructure.

## 4.6 Test and Evaluation

The concern over test and evaluation issues is nearly as old as the concept of NCO. In their book on NCO, Alberts, Garstka, and Stein discuss implications of the concept, stating that: "Testing systems will become far more complex since the focus will not be on the performance of individual systems, but on the performance of federations of systems." This leads to the conclusion that traditional engineering techniques for evaluating network-centric software architectures will not be able to completely meet the network-centric software systems test and evaluation need. Traditional techniques are likely necessary, but by no means sufficient.

We do not claim the list of challenges outlined above is comprehensive. As the software architecture community gains more insight into network-centric software systems, we believe more challenges will be identified.

# 5. AN ARCHITECTURAL APPROACH TO ADDRESSING THESE CHALLEGES

Architects think about software in different ways: in terms of system modules, in terms of components and connectors among them, or in terms of system allocation to its environment. Clements et al call these perspectives viewtypes [6]. Relevant to our discussion here is the components-and-connectors viewtype, C&C viewtype for short. This viewtype corresponds to the way architects look at the software system as a set of elements and their interactions at runtime. Within the confines of the C&C viewtype, several recurring patterns have been recognized. These patterns are called architectural styles. For instance, client-server, publish-subscribe, peer-to-peer, and others are known general-purpose architectural styles that come under the umbrella of the C&C viewtype.

Because network-centric software systems are characterized by a dynamic runtime behavior, the approach to addressing the aforementioned challenges must start at the architectural level, particularly when viewing these systems from a behavioral viewpoint. Architects design architectural views (architecture) of a particular system from one perspective at a time. They use their own experiences with styles that they know and that successfully worked for them in the past to design these views. The question is": *Do existing styles cover all what architects need to design network-centric software systems?* The answer to this question is quite simple: "No, they don't."

Architectural styles emerge as formal architectural approaches after architects have been using these styles for a while. Once a style proves to be effective in solving a particular design problem, architects then formalize its definition and documentation, and make it available as a choice in the architectural design space. Our research in the area of network-centric software systems has led us to recognize that there is a trend that is unique in the way network-centric software systems are built, and that it is different from existing styles. Thus, we propose the recognition of a new architectural style under the C&C viewtype. We argue that existing styles, individually, cannot respond to the emerging challenges of network-centric software systems; that is application and data (both new and legacy) integration. Thus, we have begun defining the elements of this new style by taking advantage of beneficial characteristics of many existing styles to describe the diverse nature of network-centric software systems.

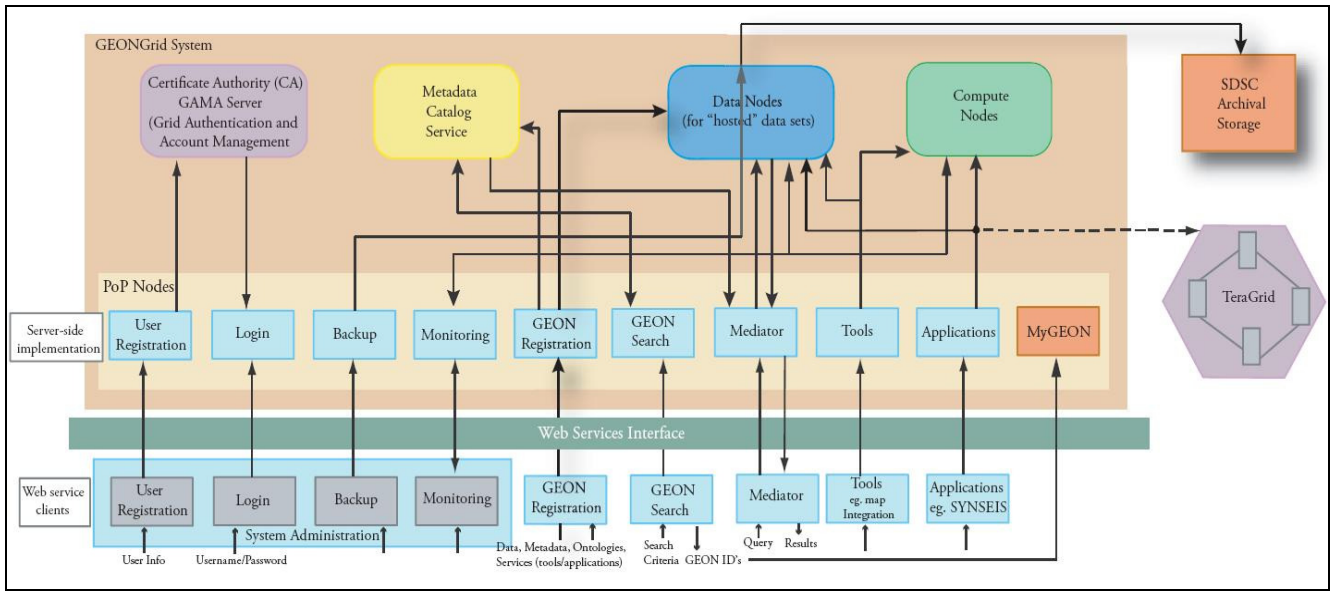

**Figure 2. The GEON software architecture**

An essential part of documenting a new style is to develop a style guide that records the specialization and constraints the style imposes on its elements and their interactions. Our current research focused on achieving this milestone and making it available to the software architecture community.

# 6. THE GOEN CYBERINFRA-STRUCTURE: AN ILLUSTRATIVE EXAMPLE OF NETWORK-CENTRICITY USEFULLNESS

The main focus of network-centric software systems is application and data integration to solve larger and complex problems. For illustration purposes, we discuss an ongoing research endeavor that we have been part of for two years (May 2004 through April 2006), which exhibits an evidence of the usefulness of network centricity in solving real life problems. In discussing this example, we demonstrate why this system fits our definition of a network-centric system.

The GEON (GEOscience Network) research project is a project that was created to respond to the pressing need in the geosciences communities to interlink and share multidisciplinary datasets to model the complex dynamics of earth systems [7].

This is a prime example of how scientists are moving towards more collaboration to tackle complex problems. Being the provider of the software part of the solution, software engineers are on the front line to face such challenging complexities.

The GEONgrid system is a network-centric computing infrastructure that manages access to remote data collections and services. It consists of a coalition of hardware nodes (*system of systems*) that are deployed with a standardized GEON software stack. The software infrastructure is based on a service-oriented architecture (SOA) using Web services and Grid Security Infrastructure (GSI) (*networked configuration*). The GEON software stack is packaged and tested as a unit to ensure that the components work together reliably, as a single software package. Currently, the GEONgrid spans principal-investigator (PI) institutions as well as a number of other collaborating sites including international sites. Figure 2 shows the underlying architecture for GEON systems.

This architecture shows how GEON is integrating existing tools, application, and databases that are dispersed worldwide amongst the 22 participating U.S. institutions and many others partners around the globe. The purpose is to create a cyberinfrastructure that enables geoscientists to engage in questions about earth that were not thought of before. This initiative makes raw rock data from an institution in Indonesia instantly available for a Virginia Tech geoscientist who is specialized in that particular rock type. This scientist can, in turn, makes analysis and share results with the rest of the community for peer reviewing (*dynamic runtime behavior*) - the interacting resources are not necessarily known at design time but they are dynamically invoked once a transaction is initiated. The result of such cooperation is a better understanding of natural earth systems. In this type of collaboration, there is a *decentralized control* over the functionality of the system. All participating institutions have equal access to all resources. The results of bringing together data and analysis tools from around the globe is intended to help in the prediction of natural disasters such as tsunamis, hurricanes, volcanoes' eruption, and others. Addressing these kinds of problems has only recently been possible. However, with the use of a network-centric approach, such endeavors are not only feasible, but also achievable.

# 7. CONCLUSION

Network-centric software systems embody the answer to the pressing need to integrate existing software assets with newly developed applications, and to the growth in size of software-intensive systems that are being developed. They also exemplify a new way of thinking about software systems. Network-centric software systems are the outcome of the inevitable shift from developing a system of statically-distributed resources to a system of dynamically-distributed components that are owned and managed by different entities, and that provide task specific services that can be used to achieve a larger goal. This shift has brought about a need for new architectural approaches to design such systems. In this paper, we have pointed out the challenges that need to be addressed concerning the software architecture of network-centric software systems. Further, we have constructed a case for the need for a new architectural style that will address these issues. We have also identified the basic principles and characteristics of network-centricity by investigating the roots of the term "network-centric" and its association with the software architecture community.

## 8. FUTURE WORK

This work is by no means complete. It is part of our ongoing research activity. Currently, we are drafting the pieces of the style elements for this new architectural style. A number of activities remain to be completed to substantiate our proposed style. Amongst these activities is to produce a complete style guide with examples and scenarios describing when and how to use this style. Also, much work needs to be done in terms of comparing the network-centric style with existing ones to validate its standalone status. Distinguishing this style from existing ones will help architects understand and adopt this style in their practice more easily.

## 9. ACKNOWLEDGEMENTS

We would like to acknowledge Shawn Bohner, a member of my advisory committee, for his contribution to the research presented in this paper.